# Research on the spectral reconstruction of a low-dimensional filter array micro-spectrometer based on a truncated singular value decomposition-convex optimization algorithm

Jiakun Zhang,[1,2] Liu Zhang,[1,2] Ying Song,[1,2] Yan Zheng.[1,2] *

[1] College of Instrumentation and Electrical Engineering, Jilin University, Changchun, Jilin 130012, China

2 National Engineering Research Center of Geophysics Exploration Instruments, Jilin University, Changchun 130061, China

*zhengyan19@mails.jlu.edu.cn.

**Abstract:** Currently, the engineering of miniature spectrometers mainly faces three problems: the mismatch between the number of filters at the front end of the detector and the spectral reconstruction accuracy; the lack of a stable spectral reconstruction algorithm; and the lack of a spectral reconstruction evaluation method suitable for engineering. Therefore, based on 20 sets of filters, this paper classifies and optimizes the filter array by the K-means algorithm and particle swarm algorithm, and obtains the optimal filter combination under different matrix dimensions. Then, the truncated singular value decomposition-convex optimization algorithm is used for high-precision spectral reconstruction, and the detailed spectral reconstruction process of two typical target spectra is described. In terms of spectral evaluation, due to the strong randomness of the target detected during the working process of the spectrometer, the standard value of the target spectrum cannot be obtained. Therefore, for the first time, we adopt the method of joint cross-validation of multiple sets of data for spectral evaluation. The results show that when the random error of +/− 2 code values is applied multiple times for reconstruction, the spectral angle cosine value between the reconstructed curves becomes more than 0.995, which proves that the spectral reconstruction under this algorithm has high stability. At the same time, the spectral angle cosine value of the spectral reconstruction curve and the standard curve can reach above 0.99, meaning that it realizes a high-precision spectral reconstruction effect. A high-precision spectral reconstruction algorithm based on truncated singular value-convex optimization, which is suitable for engineering applications, is established in this paper, providing important scientific research value for the engineering application of micro-spectrometers.

**Keywords**

# Introduction

As an important instrument for obtaining spectral information, spectrometers are gradually developing toward integration and miniaturization. Currently, the engineering of miniature spectrometers mainly faces three problems. The first is the mismatch between the number of filters at the front end of the detector and the spectral reconstruction accuracy; the second is the lack of a stable spectral reconstruction algorithm; and the third is the lack of a spectral reconstruction evaluation method suitable for engineering. In recent years, micro-spectrometers based on filter arrays have received extensive attention [1–5]. The advantage of a miniature spectrometer over a traditional spectrometer is that it does not need prisms, gratings, or other optical elements to split light [6, 7], but through the mathematical relationship between target spectrum, filter array, detector quantum efficiency, and gray value, $AX = B$ is calculated.

The spectral reconstruction algorithm is the core of the miniature spectrometer, and its essence is solving the equation system $AX = B$ with high precision. Scholars engaged in related research have proposed many high-precision spectral reconstruction algorithms. However, these algorithms are often proposed under ideal conditions without considering the errors generated during the development and operation of the spectrometer. Currently, there are three main problems in the miniaturization of micro-spectrometers. First, regarding the issue of the number of filter arrays, since the increase in the number of filters will lead to high spectral reconstruction accuracy, the number of filters used in previous related studies has exceeded 190 [8, 9]. However, due to the limited area of the photosensitive surface of the spectrometer detector, it is impossible to have hundreds of filters; typically, only about 20 filters can be used. Second, in the problem of a universal high-precision spectral reconstruction algorithm, the spectrometer will produce many errors in the working process; for example, the processing error of the filter array, the noise of the detector itself, and the error of the gray value of the detector. Related research has not explained the error source and measurement error range, and instead has simply imposed a random error of 20 dB; thus, it could not simulate the real error environment or analyze the error of the filter. Finally, regarding the issue of the engineered spectral reconstruction evaluation criteria, the evaluations of spectral reconstructions in previous studies were all performed under the assumption of known targets. In practice, the objects observed in the working process of the miniature spectrometer have strong randomness, and we cannot obtain the true curve of the spectrum in advance. Therefore, it is necessary to perform a reconstruction evaluation on the reconstructed spectrum under the condition of an unknown target, and thus a high-precision and stable spectral reconstruction algorithm suitable for engineering is required. Based on the above three situations, this paper first selects the filter array based on the K-means and particle swarm algorithms, then uses the truncated singular value decomposition-convex optimization (Cvx) algorithm to reconstruct the spectrum, and finally conducts a spectral evaluation according to the multi-group cross-validation method, which satisfies the engineering requirements. Therefore, for the first time, we adopt the method of joint cross-validation of multiple sets of data for spectral evaluation. The results show that when the random error of +/− 2 code values is applied multiple times for reconstruction, the spectral angle cosine value between the reconstructed curves becomes more than 0.995, which proves that the spectral reconstruction under this algorithm has high stability. At the same time, the spectral angle cosine value of the spectral reconstruction curve and the standard curve can reach above 0.99, thus realizing a high-precision spectral reconstruction effect. A high-precision truncated singular value decomposition-Cvx spectral reconstruction algorithm suitable for engineering applications is established in this paper, which provides important scientific research value for the engineering application of micro-spectrometers.

## Working principle of miniature spectrometer

The working principle of the spectrometer is shown in Fig. (1). The target spectrum $X(\lambda)$ passes through the filter array at the front of the detector $W(\lambda)$. The gray value B [10–12] is obtained after modulation of the quantum efficiency $G(\lambda)$ of the detector.

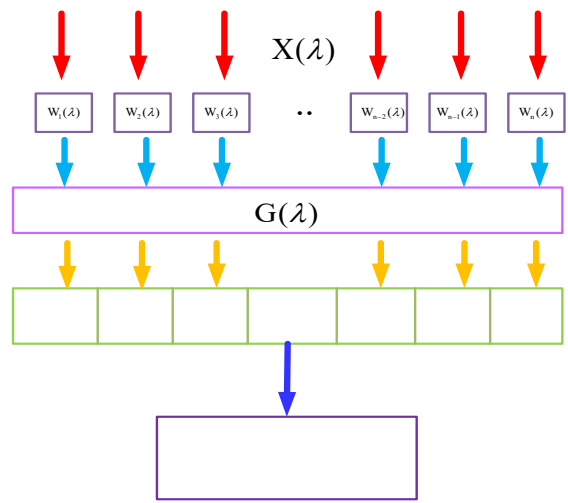

**Figure 1.** Schematic of proposed micro-spectrometer.

The relationship among $X(\lambda)$, $W(\lambda)$, and $G(\lambda)$ is shown in Eq. (1).

$$\int_{\lambda_2}^{\lambda_1} W(\lambda) \cdot G(\lambda) \cdot X(\lambda) = B \quad (1)$$

Let $A(\lambda) = W(\lambda) \cdot G(\lambda)$, then Eq. (1) is simplified as Eq. (2).

$$\int_{\lambda_2}^{\lambda_1} A(\lambda) \cdot X(\lambda) = B \quad (2)$$

After discretizing Eq. (2), we can obtain Eq. (3).

$$A \cdot X = B \quad (3)$$

The specific expression of Eq. (3) is shown in Eq. (4).

$$\begin{pmatrix} a_{11} & \cdots & a_{1n} \\ \vdots & \ddots & \vdots \\ a_{n1} & \cdots & a_{nn} \end{pmatrix} \cdot \begin{pmatrix} x_1 \\ \vdots \\ x_n \end{pmatrix} = \begin{pmatrix} b_1 \\ \vdots \\ b_n \end{pmatrix} \quad (4)$$

The spectral reconstruction problem of the miniature spectrometer thus becomes a problem of solving equations. The micro-spectrometer filter array is shown in Fig. 2(a). The number of filter arrays on the front end of the detector is about 20. The spectral range of the micro-spectrometer for visible light is 400–900 nm, and thus it is necessary to use 20 filters to calculate the spectral information of the target spectrum from 400 to 900 nm.

However, the working process of the spectrometer will be affected by a variety of error sources. These error sources are divided into four categories: The first is the influence of the noise of the detector itself; the second is the influence of the optical system aberration; the third is the influence of stray light on the energy of the detector; and the fourth is the filter-processing error. The first three errors have a greater impact on the detector gray value (matrix **B**), and the fourth error will cause errors in the **A** matrix. After detection, the error experienced by the detector is about two gray values as shown in Fig. 2(b): This is far beyond the 80 dB error of the signal-to-noise ratio in the previous study.

The methods used for Eq. (3) in previous related studies include GPSR [13], OMP [14], and the CNN algorithm [15]; however, these methods all have the same drawbacks. First, the number of filters required is large. The second is that they must be solved under the condition of a known target. The third is that the applied error is small in the simulation process, which does not meet the actual work requirements of the spectrometer.

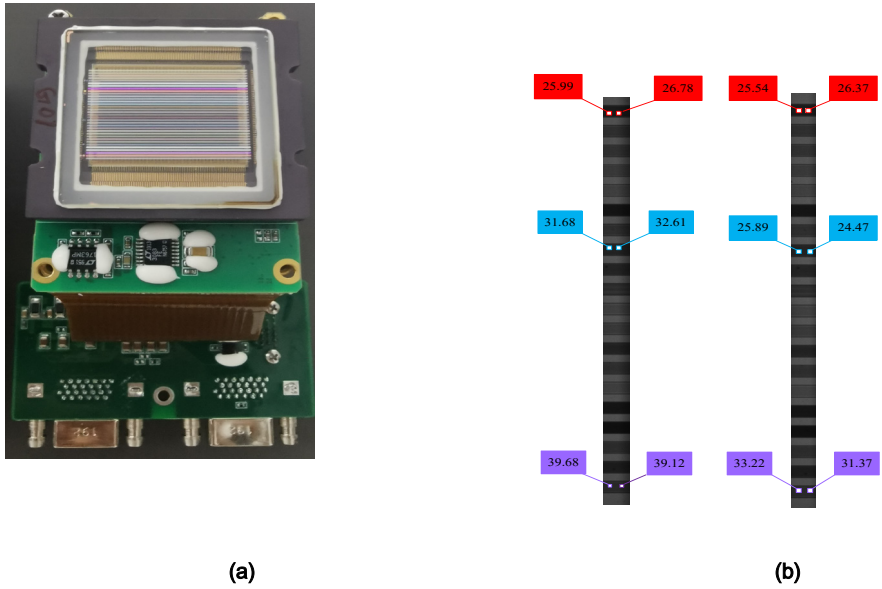

**Figure 2.** (a) Schematic diagram of the micro-spectrometer filter array; (b) grayscale image.

## Filter optimization selection process

### *The principle of filter selection*

The analysis in the second section indicates that there is a large error in the working process of the spectrometer, which has a huge impact on the solution accuracy of Eq. (3). In fact, the excessively large size of matrix **A** in Eq. (3) is the root cause of poor solution accuracy. However, the precise solution of Eq. (3) can be improved through different filter combinations. The essence of filter combination screening is to reduce the condition number of matrix **A** under the condition of full rank of matrix **A**. The selection of filter array combinations is based on Eq. (5) [16, 17] and Eq. (6) [18]. Eq. (5) calculates the cosine of the angle between the two curves, and its purpose is to judge the similarity of the two curves. A larger $cos\theta$ value implies that the curves are more similar. The purpose of Eq. (5) is to ensure the full rank of the **A** matrix and the uniqueness of the solution of Eq. (3). Terms $x_i$ and $y_i$ represent discrete points at wavelengths of the two spectral curves. Eq. (6) is the definition of matrix condition number, and the condition number can be determined by the ratio of the largest eigenvalue ($\sigma_{max}$) to the smallest eigenvalue ($\sigma_{min}$). A larger condition number leads to a more ill-conditioned matrix.

$$\cos\theta = \frac{\sum_{i=1}^{n} x_i y_i}{\sqrt{\sum_{i=1}^{n} x_i^2} \cdot \sqrt{\sum_{i=1}^{n} y_i^2}} \quad (5)$$

$$Cond(A) = \frac{\sigma_{max}}{\sigma_{min}} \quad (6)$$

*K-means filter clustering and optimization*

First, 20 $A(\lambda)$ are numbered including the multiplication of the detector quantum efficiency curve and the filter transmittance curve (20 $A(\lambda)$ curves are given in the Appendix). The K-means algorithm was used to cluster 20 filters based on Eq. (5). In each cluster, the similarity between $A(\lambda)$ is high, and the similarity in different $A(\lambda)$ clusters is low. For example, Fig. 3 shows the clustering result with 10 clusters; each circle represents the filter number.

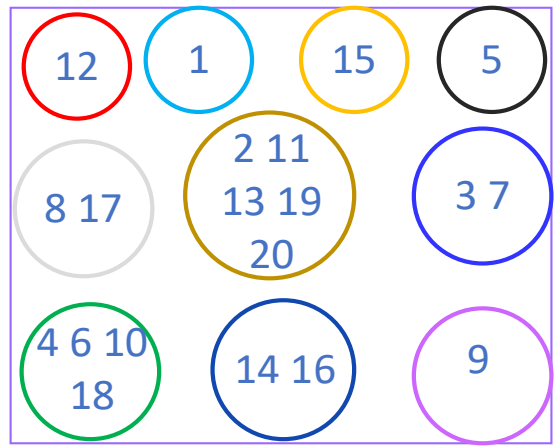

**Figure 3.** K-means clustering results when number of clusters is 10.

After the clustering is completed, Eq. (6) is used as the evaluation function, and the particle swarm optimization algorithm is used for optimization to find the minimum condition number. The optimization results are shown in Table 1.

Table 1. Filter combination optimization results.

| Number of clusters | Optimal filter combination | Condition number | Full rank |
| --- | --- | --- | --- |
| 5 | 14 8 4 12 5 | 43.82 | Yes |
| 6 | 9 6 3 5 15 14 | 62.49 | Yes |
| 7 | 1 15 13 11 6 7 9 | 102.06 | Yes |
| 8 | 11 20 7 9 14 8 12 16 | 144.01 | Yes |
| 9 | 8 4 17 20 6 16 12 1 5 | 162.41 | Yes |
| 10 | 12 1 15 5 8 19 3 10 16 9 | 2007.40 | Yes |
| 12 | 12 14 15 6 5 10 8 7 4 9 12 1 | 13415.29 | Yes |
| 14 | 16 13 5 12 8 20 2 1 14 17 4 10 15 9 | 209843.32 | Yes |
| 15 | 5 4 13 8 20 10 17 7 9 18 19 11 1 14 15 | 1001439 | Yes |

Table 1 shows that the matrices are all full rank after the filter is selected, and the matrix condition number becomes larger and larger an increase in the number of clusters. In turn, the ill-conditioned degree of the matrix becomes higher and higher, which is very unfavorable for an Eq. (3) solution. It is necessary to make full use of the optimization results in Table 1 for spectral reconstruction on the basis of filter combinations with smaller condition numbers.

## Spectral reconstruction algorithms

### Truncated singular value decomposition (Tsvd)

Truncated singular value decomposition [19-20] is a classic method for dealing with large condition number matrices. After the matrix undergoes singular value decomposition, a set of eigenvalue matrices arranged from large to small are generated as shown in Eq. (7) and Eq. (8) ($\sigma_1 > \sigma_2 > \ldots > \sigma_n$). The minimum value of these eigenvalues is often very small, and these smaller eigenvalues are the main factors affecting the larger condition number of the matrix. Relatively large eigenvalues represent more reliable parts, and smaller eigenvalues represent large floating and unreliable parts. Therefore, our most direct approach is to "truncate" the eigenvalue matrix and discard the smaller part of the eigenvalue matrix to reduce the condition number of the matrix.

$$A = UDV^T \qquad (7)$$

$$D = diag(\sigma_1, \sigma_2, \cdots, \sigma_n) \quad (8)$$

When the eigenvalues are not "truncated," Eq. (3) is shown as Eq. (9).

$$X = \sum_{i=1}^{n} \sigma_i^{-1} v_i u_i^T B \quad (9)$$

However, in order to make the solution more accurate, this paper first improves **D** according to Eq. (10), which will somewhat reduce the condition number of matrix **A**.

$$D = D + \frac{min(D)}{D} \quad (10)$$

When "truncation" is performed on the eigenvalues, the first t eigenvalues are retained, and t is also called the "truncation threshold." The solution of Eq. (3) at this time is shown in Eq. (11).

$$\mathbf{X} = \sum_{i=1}^{t} \sigma_i^{-1} v_i u_i^T \mathbf{B} \quad (11)$$

*Convex optimization*

Eq. (3) is a typical convex problem. Cvx [21–23] is a common method for solving Eq. (3). The general form of Cvx is shown in Eq. (12).

$$\begin{aligned} &\text{minimize } f_0(x) \\ &\text{subject to } f_i(x) \leq 0,\ i=1,2,\cdots,m \quad (12) \\ &\qquad\qquad h_i(x) = 0,\ i=1,2,\cdots,n \end{aligned}$$

In Eq. (12), $f_0(x)$ is the objective function, $f_i(x)$ is the inequality constraint, and $h_i(x)$ is the equality constraint. The Cvx problem becomes a linear programming problem when the objective function and constraint function of Cvx are both affine functions. Eq. (12) indicates that the Cvx can obtain more accurate results with appropriate constraints, but the constraints are very important to the solution.

*Spectral reconstruction based on Tsvd and Cvx*

Although Tsvd and Cvx can each obtain relatively accurate results, they each also have their own shortcomings. Tsvd has two disadvantages: The first is that the discrete values solved may have negative values, and the second is that when the condition number is too large, too much spectral information needs to be lost to obtain a lower condition number. The solution method of Cvx is more flexible, and different solutions can be obtained by adding different constraints. However, the constraints are critical to obtaining an accurate solution. If the constraints are insufficient, then discrete values with large deviations from the standard value will be obtained. In order to obtain a high-precision reconstruction curve, this paper combines the two algorithms, and takes the result obtained by the Tsvd algorithm as the constraint of the Cvx algorithm to reconstruct two typical spectral curves.

First, an error of three thousandths is applied to matrix **A**, and a random error of plus or minus three gray values is applied to matrix **B**. According to the results obtained from the filter combination optimization in Table 1, the discrete values with the cluster numbers 5, 6, and 7 are first solved by truncated singular value decomposition.

Table 2. Number of clusters is 5 (target 1).

| Discrete values | Standard value | Error |
|---|---|---|
| 0.4280 | 0.4603 | −0.0323 |
| 0.4866 | 0.5275 | −0.0409 |
| 0.5204 | 0.4792 | 0.4127 |
| 0.5275 | 0.5088 | 0.0187 |
| 0.5275 | 0.5284 | −0.0009 |

Table 3. Number of clusters is 6 (target 1).

| Discrete values | Standard value | Error |
|---|---|---|
| 0.4099 | 0.4497 | −0.0397 |
| 0.5460 | 0.5295 | 0.0165 |
| 0.4565 | 0.4967 | −0.0402 |
| 0.5007 | 0.4799 | 0.0208 |
| 0.5610 | 0.5220 | 0.0390 |
| 0.5551 | 0.5273 | 0.2784 |

Table 4. Number of clusters is 7 (target 1).

| Discrete values | Standard value | Error |
|---|---|---|
| 0.4806 | 0.4422 | 0.0384 |
| 0.5390 | 0.5224 | 0.0167 |
| 0.4700 | 0.5175 | −0.0475 |
| 0.4627 | 0.4764 | −0.0137 |
| 0.5364 | 0.4911 | 0.0452 |
| 0.5680 | 0.5305 | 0.0375 |
| 0.5003 | 0.5257 | −0.0255 |

Tables 2–4 show that the number of discrete points is small although the three sets of results calculated by truncating singular values have high precision. Taking the calculation result of the cluster class as 5, the wavelengths corresponding to these five discrete points are 450 nm, 550 nm, 650 nm, 750 nm, and 850 nm. When fitting these five points in the range of 400–900 nm, there may be points that are quite different from the standard values because there are no discrete points in the range of 400–450 nm and 850–900 nm. Thus, upon adding two discrete points of 400 nm and 425 nm in the range of 400–450 nm, the value of these two discrete points is equal to the discrete value at 450 nm. Similarly, upon adding two discrete points of 850 nm and 900 nm in the range of 850–900 nm, these two discrete points have a value equal to the discrete value at 875 nm. The fitting results are shown in Fig. 4.

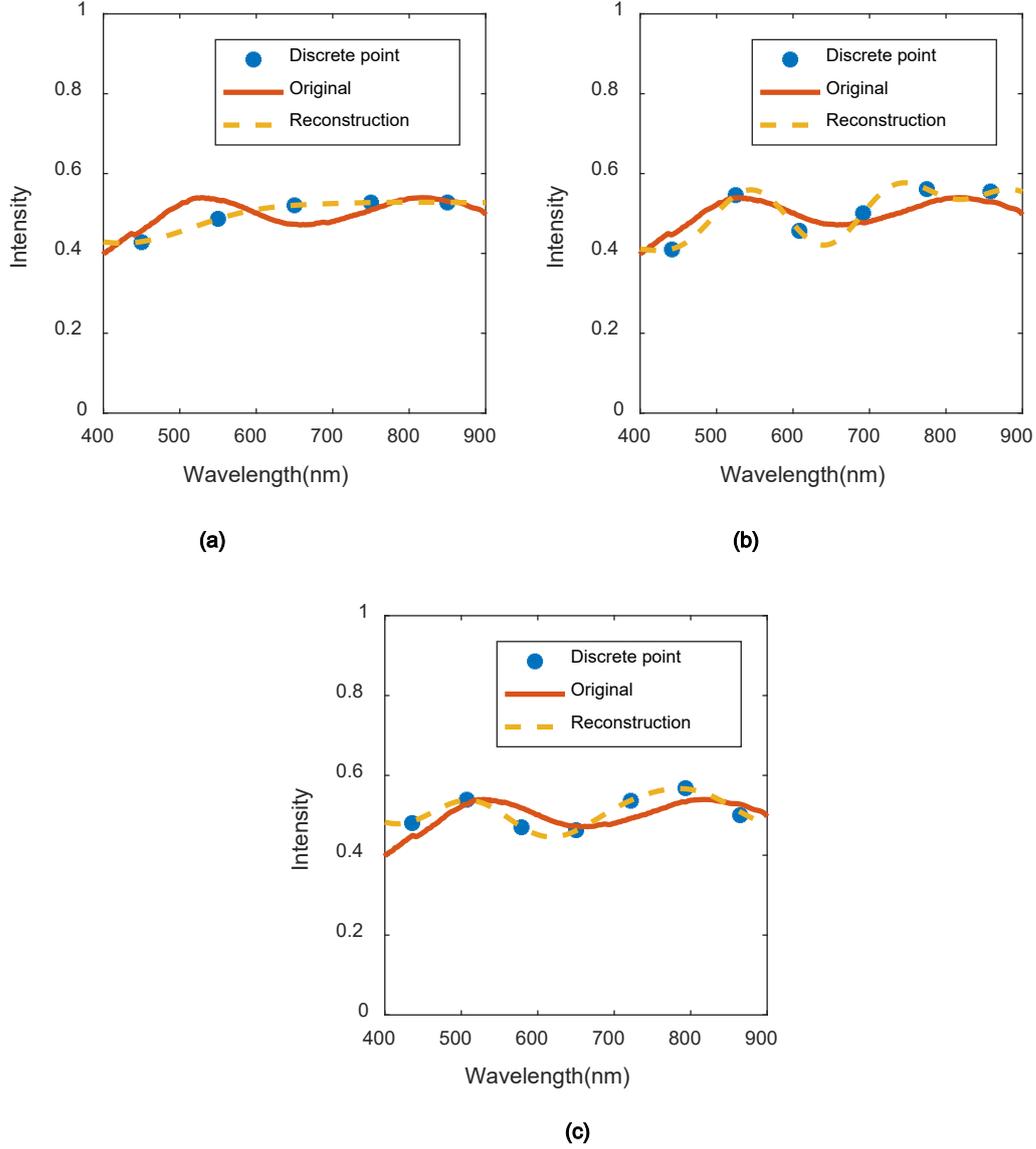

**Figure 4.** Calculation results of Tsvd: (a) number of clusters is 5, (b) number of clusters is 6, and (c) number of clusters is 7.

The above three results are added to the Cvx as constraints, and the Cvx expression is shown in Eq. (13).

$$\begin{aligned}
&\text{minimize } ||\mathbf{AX}\text{-}\mathbf{B}||^2 \\
&\text{subject to } |X_i\text{-}X_{i+1}|<R,\ i=1,2,\cdots,n-1 \\
&\qquad |\frac{X_i+X_{i+1}}{2}\text{-}Y_j|<T \qquad i=1,3,\cdots,n\ ;\ j=1,2,\cdots\frac{n}{2} \\
&\qquad X_i < max(Y) \\
&\qquad X_i > min(Y)
\end{aligned} \qquad (13)$$

In Eq. (13), $|X_i\text{-}X_{i+1}|<R$ is the constraint of two adjacent discrete points, $|X_i\text{-}mean|<T$ is the constraint between each discrete point and the target mean, $|\frac{X_i+X_{i+1}}{2}\text{-}Y_j|<T$ is the constraint of truncated singular value decomposition, and $Y_j$ is the result of truncated singular value calculation.

Terms $X_i < max(Y)$ and $X_i > min(Y)$ are boundary constraints, which determine the solution range of CVX. Solving discrete values and fitting spectral curves are shown in Fig. 5.

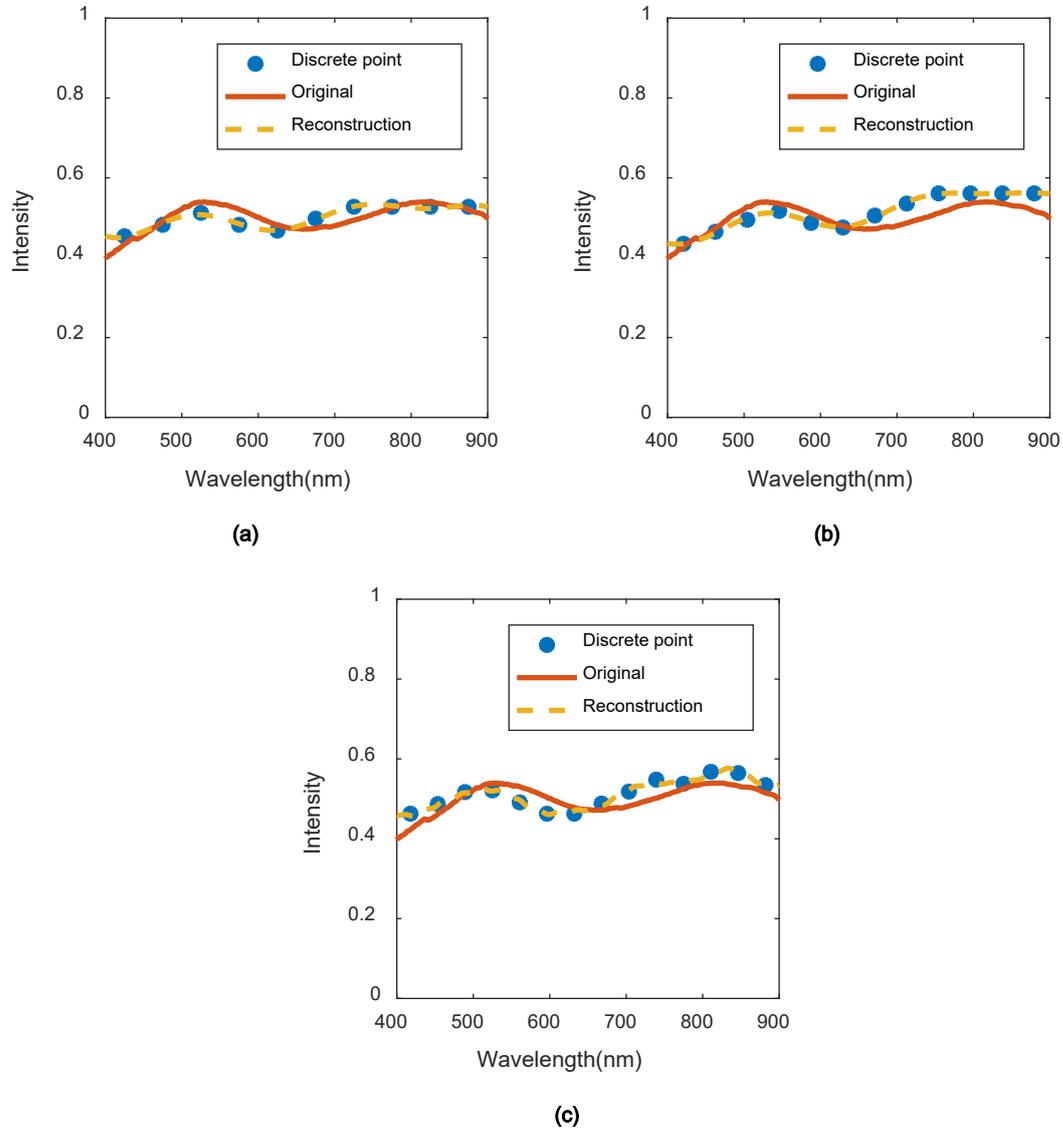

**Figure 5.** Calculation results of Tsvd constrained Cvx, (a) number of clusters is 10, (b) number of clusters is 12, and (c) number of clusters is 14.

The above simulation is aimed at the reconstruction result of the relatively flat spectral curve. To verify the universality of the algorithm, the reconstruction is performed on the spectral curve with obvious peaks (target 2). Similarly, we first applied an error of three thousandths to matrix **A**, and then applied a random error of plus or minus 3 grayscale values to matrix **B**. The truncated singular value decomposition results are shown in Tables 5–7.

Table 5. Number of clusters is 5 (target 2).

| Discrete values | Standard value | Error |
|---|---|---|
| 0.3728 | 0.3207 | 0.0528 |
| 0.3401 | 0.3937 | −0.0536 |
| 0.2191 | 0.2569 | −0.0377 |
| 0.1976 | 0.1976 | 0 |
| 0.2570 | 0.2113 | 0.4566 |

Table 6. Number of clusters is 6 (target 2).

| Discrete values | Standard value | Error |
|---|---|---|
| 0.2049 | 0.3089 | −0.1040 |
| 0.4485 | 0.3972 | 0.0513 |
| 0.2675 | 0.3299 | −0.0624 |
| 0.2740 | 0.2087 | 0.0652 |
| 0.2691 | 0.2010 | 0.0681 |
| 0.1935 | 0.2118 | −0.0182 |

Table 7. Number of clusters is 7 (target 2).

| Discrete values | Standard value | Error |
|---|---|---|
| 0.3193 | 0.3009 | 0.0184 |
| 0.4172 | 0.3880 | 0.0292 |
| 0.3287 | 0.3786 | −0.0500 |
| 0.2842 | 0.2548 | 0.0295 |
| 0.2633 | 0.1959 | 0.0674 |
| 0.1639 | 0.2046 | −0.0408 |
| 0.1601 | 0.2120 | −0.0519 |

Upon fitting the discrete points in Tables 5–7, we get the results in Fig. 6.

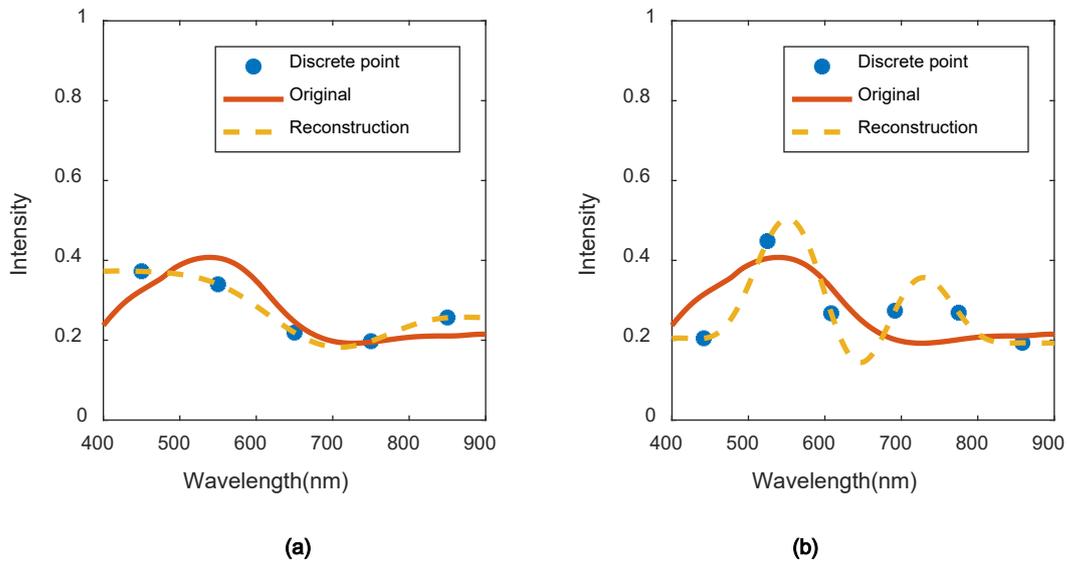

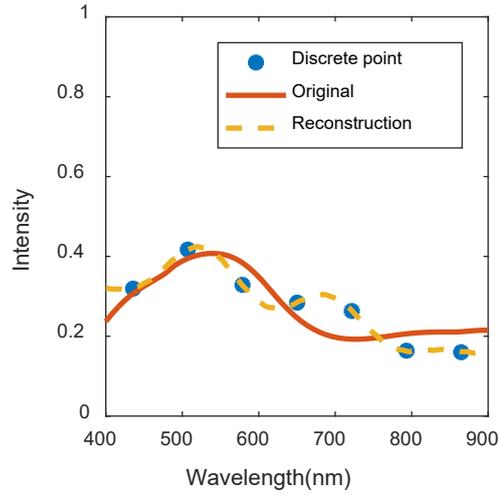

(c)

**Figure 6.** Calculation results of Tsvd: (a) number of clusters is 5, (b) number of clusters is 6, and (c) number of clusters is 7.

The Cvx solution is performed upon taking the Tsvd result as a constraint (Fig. 7).

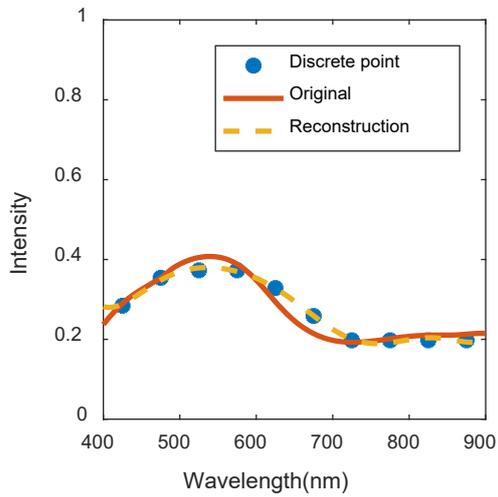

(a)

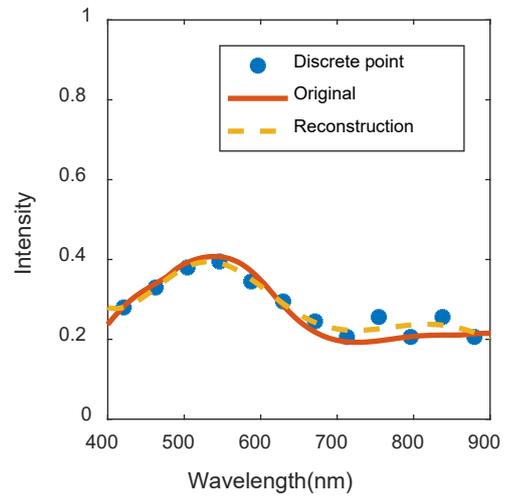

(b)

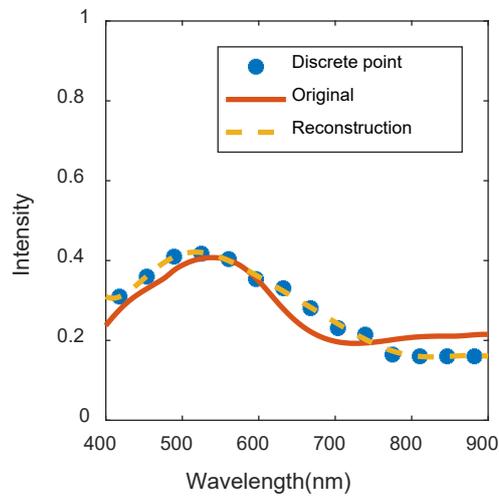

(c)

**Figure 7.** Calculation results of Tsvd constrained Cvx: (a) number of clusters is 10, (b) number of clusters is 12, and (c) number of clusters is 14.

## Spectral curve evaluation

The previously studied evaluation functions include MSE, ARE, and RE. The specific table expressions are shown in Eqs. (14), (15), and (16); here, $y_i$ is the standard value, and $\hat{y}$ is the reconstructed value.

$$ARE = \frac{\|y_i - \hat{y}\|_2^2}{\|y_i\|_2^2} \quad (14)$$

$$RE = \frac{\|y_i - \hat{y}\|_2}{\|y_i\|_2} \quad (15)$$

$$MSE = \frac{1}{n}\sum_{k=1}^{n}(y_i - \hat{y})^2 \quad (16)$$

These evaluation methods are established under the condition of a known target spectrum. The target of the spectrometer is random in the working process, and the standard value of the target spectrum is unknown; thus, the above evaluation criteria are not applicable as the evaluation criteria for the reconstruction accuracy of the target spectrum. This paper proposes a cross-validation method to solve this problem. The three same target spectral curves obtained by fitting the cluster numbers of 10, 12, and 14 were evaluated with each other, and the spectral angle cosine value ($\cos\theta$) between the three curves was calculated. The calculation results are shown in Table 8.

**Table 8.** Calculation results of $\cos\theta$ the reconstructed curve (target 1).

| Number of clusters | $\cos\theta$ |
|---|---|
| 10 and 14 | 0.9985 |
| 10 and 12 | 0.9917 |
| 12 and 14 | 0.9963 |

**Table 9.** Calculation results of $\cos\theta$ the reconstructed curve (target 2).

| Number of clusters | $\cos\theta$ |
|---|---|
| 10 and 12 | 0.9968 |
| 10 and 14 | 0.9959 |
| 12 and 14 | 0.9891 |

Table 8 shows that $\cos\theta$ values between the three reconstruction curves of target 1 are very close, and the final reconstruction curve can be obtained by averaging the three reconstruction curves in Figure 4 as shown in Fig. 8(a).

Table 9 shows that the $\cos\theta$ values of the three groups are significantly different; thus, the reconstruction curve can be obtained by averaging the combination with the highest spectral angle cosine value as shown in Fig. 8(b).

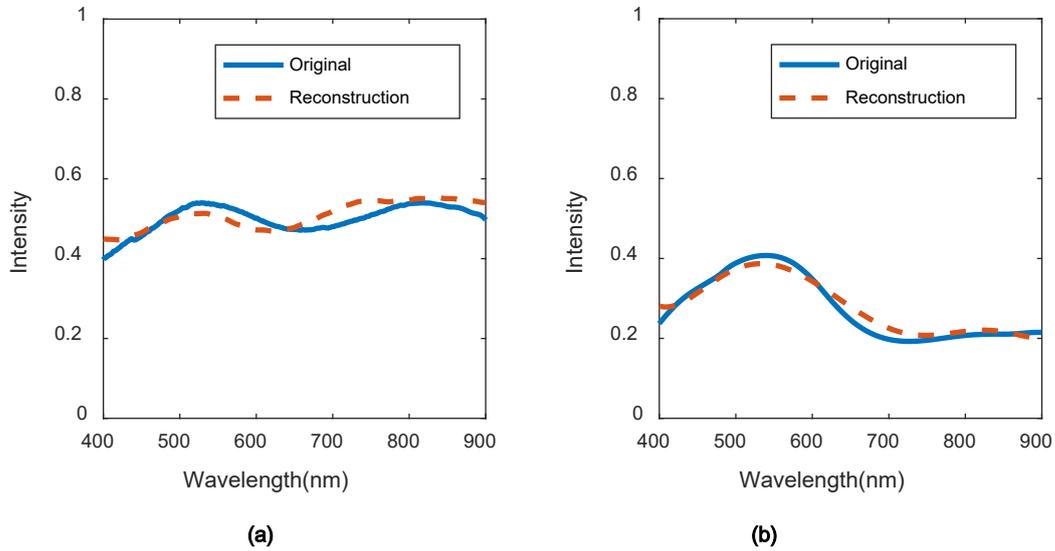

**Figure 8.** Spectral reconstruction results: (a) target 1 and (b) target 2.

## Stability verification

To verify the stability of the algorithm, 10 random errors are applied to Eq. 3, and the above steps are repeated to reconstruct the spectrum for stability verification of the reconstructed curve, as shown in Fig. 9.

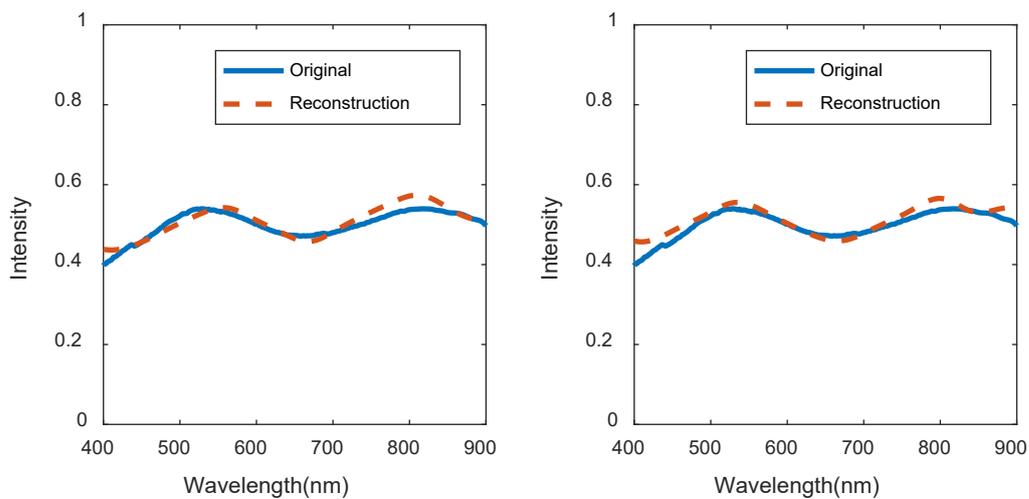

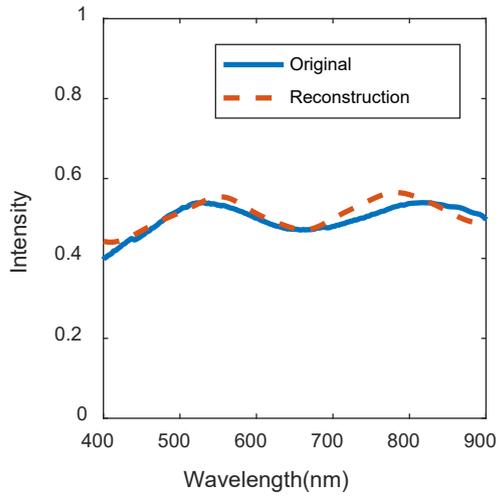 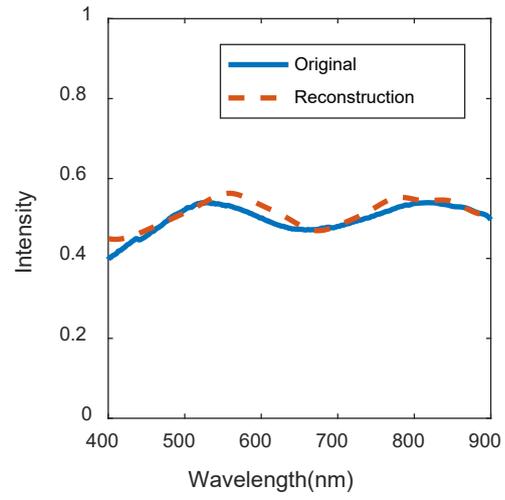

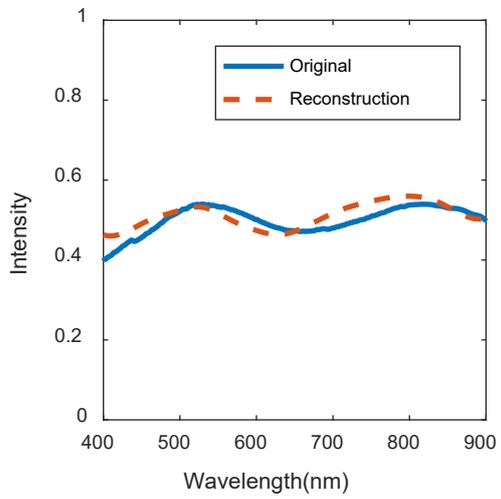 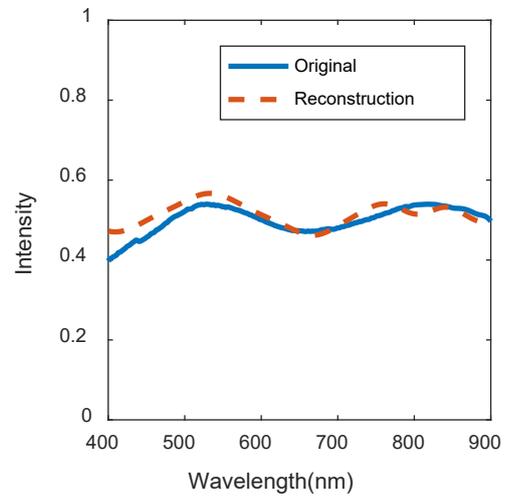

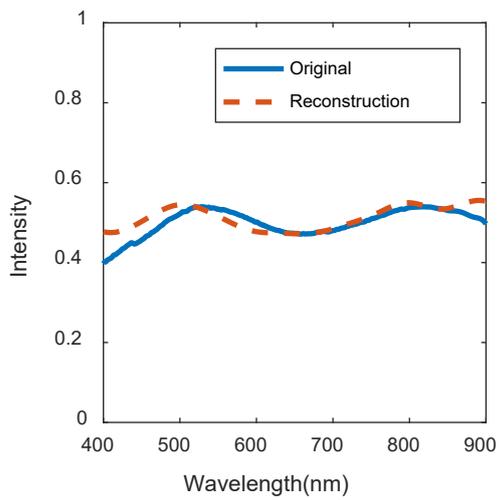 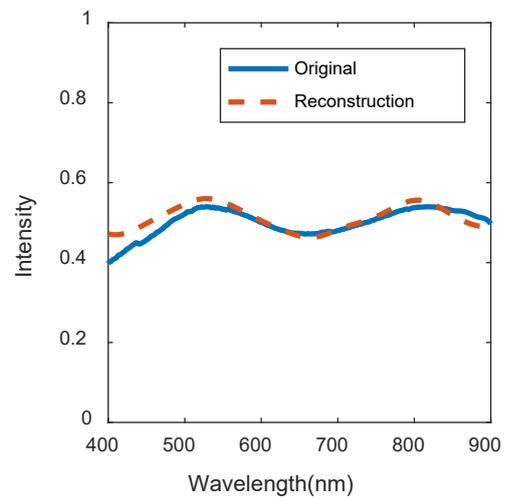

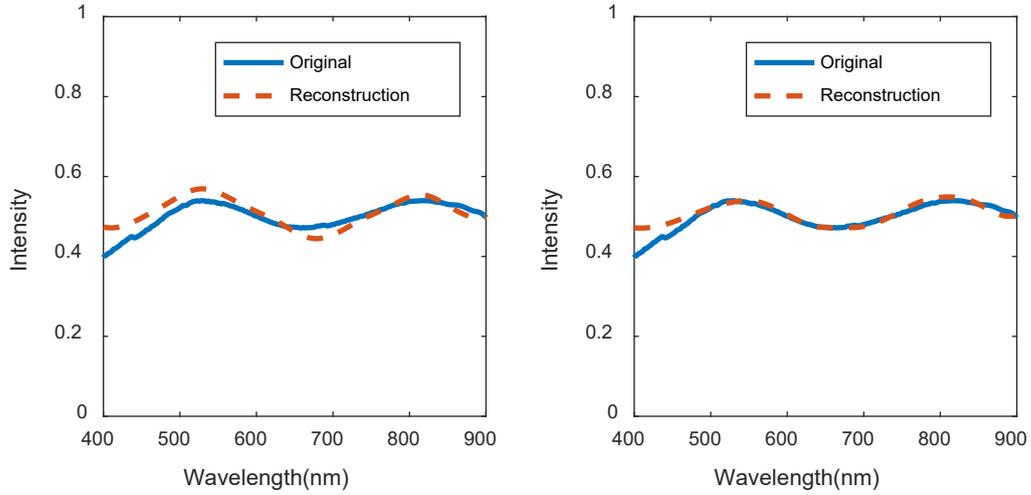

**Figure 9.** Reconstructed curve after adding 10 random errors (target 1).

The 10 reconstructed curves are now matched with the reconstructed curves in Fig. 8(a) (cross-validation), and the $\cos\theta$ values are obtained as shown in Table 10. It can be seen from Table 10 that the $\cos\theta$ values of the reconstructed curves in Figs. 8(a) and 9 reach above 0.995. It can achieve a better reconstruction effect and has strong stability.

**Table 10.** Calculated $\cos\theta$ values for the reconstructed curve (target 1).

| Number | $\cos\theta$ |
| --- | --- |
| 1 | 0.9988 |
| 2 | 0.9987 |
| 3 | 0.9986 |
| 4 | 0.9983 |
| 5 | 0.9995 |
| 6 | 0.9978 |
| 7 | 0.9990 |
| 8 | 0.9979 |
| 9 | 0.9970 |
| 10 | 0.9986 |

In the same way as the above operation, 10 reconstructed curves about target 2 are obtained, as shown in Fig. 10. The $\cos\theta$ values of the curves in Figs. 8(b) and 10 are shown in Table 11. All of the values reach above 0.992. The results show that the Tsvd-Cvx algorithm proposed in this paper can also achieve a better reconstruction effect for a target spectral curve with obvious convexity.

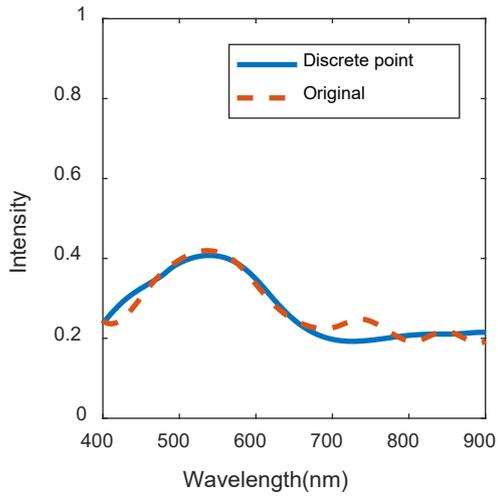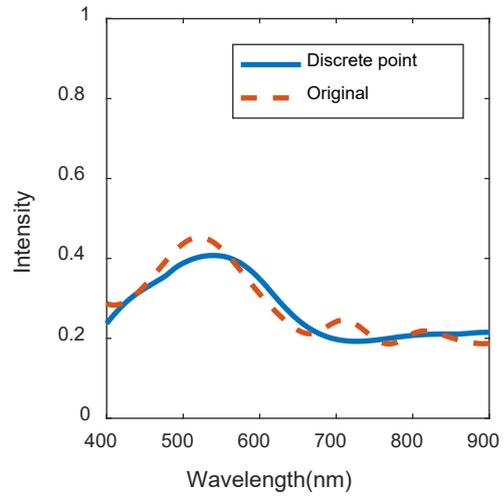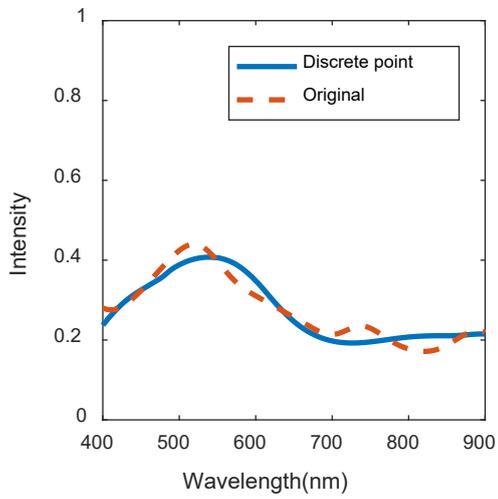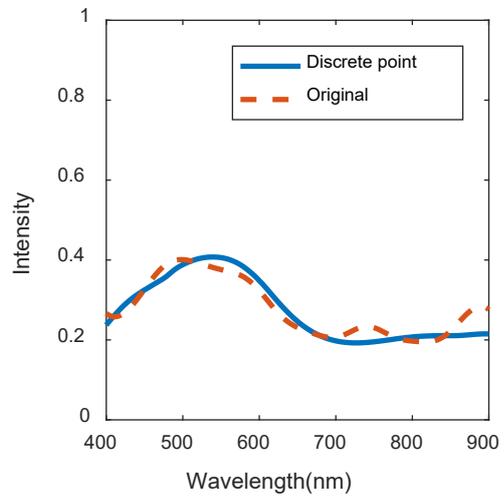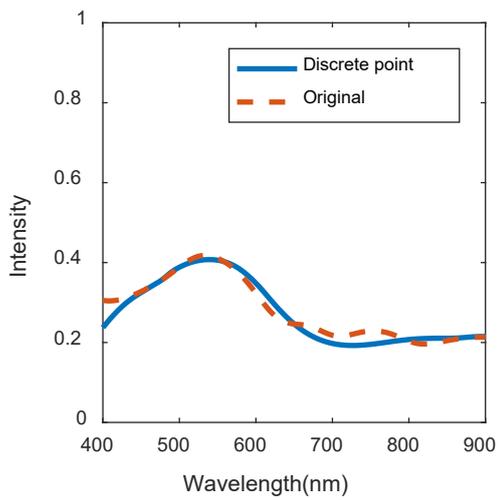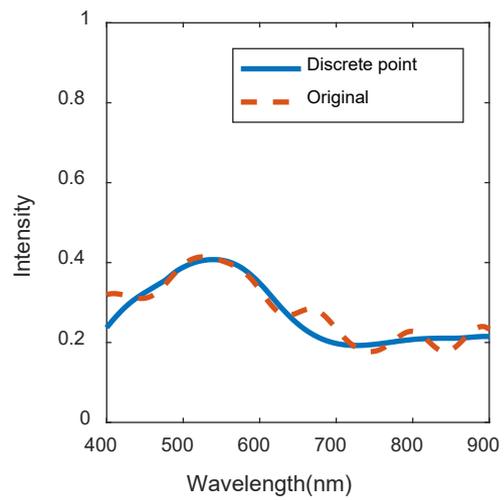

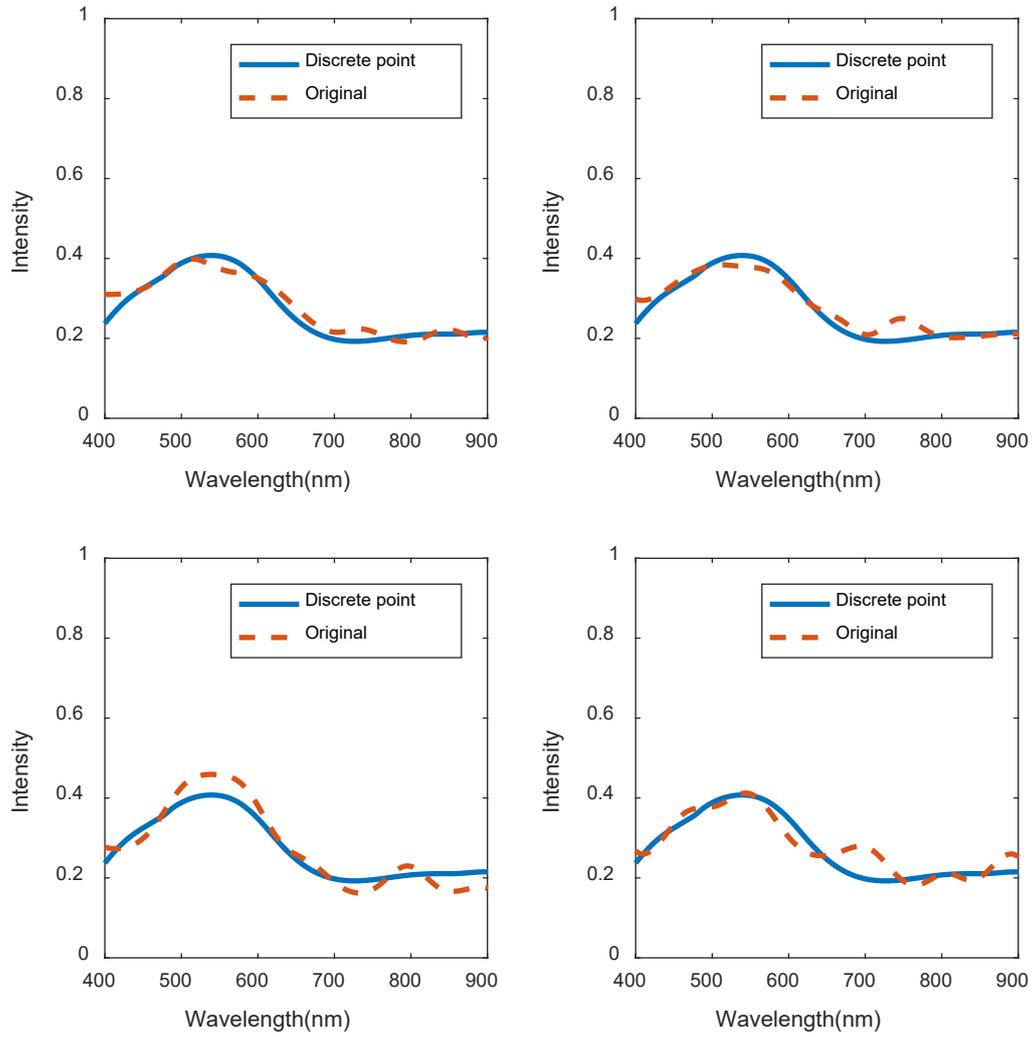

**Figure 10.** Reconstructed curve after adding 10 random errors (target 2).

**Table 10.** Calculated $\cos\theta$ values for the reconstructed curve (target 2).

| Number | $\cos\theta$ |
|---|---|
| 1 | 0.9962 |
| 2 | 0.9944 |
| 3 | 0.9963 |
| 4 | 0.9929 |
| 5 | 0.9974 |
| 6 | 0.9971 |
| 7 | 0.9991 |
| 8 | 0.9979 |
| 9 | 0.9929 |
| 10 | 0.9951 |

## Conclusion

This paper analyzes three problems in the process of micro-spectrometer engineering. First, due to the limited area of the photosensitive surface of the micro-spectrometer detector, this paper uses the K-means algorithm and particle swarm optimization to select the filter array based on 20 filter arrays. This approach reduces the condition number and satisfies the engineering requirements. Second, based on the combination of low condition number filters, this paper uses truncated singular value decomposition to solve the initial value and uses the solved initial value as a constraint of Cvx to solve the discrete value and fit. This paper used two target spectra as an example. Array A imposed a random error within three thousandths, and array B imposed a random error of plus or minus three gray values. Under these conditions, the spectrum is reconstructed. Third, this paper proposes a new evaluation method of spectral reconstruction curve, which uses multiple sets of reconstruction curves for cross-validation instead of relying on standard spectral curves to evaluate the spectrum. This leads to a high-precision reconstruction and meets engineering requirements.

## Appendix

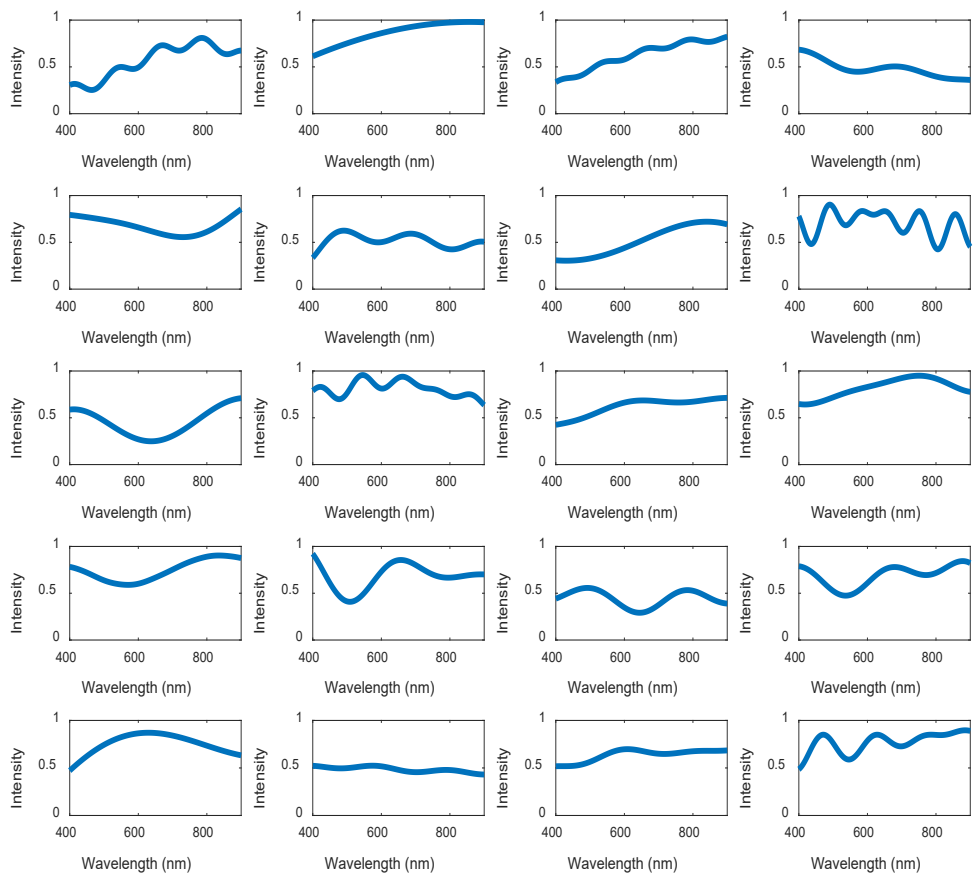

Figure 11. Twenty $A(\lambda)$ curves.

**Funding.** This study was supported by Foundation of Equipment Pre-research Area (6B2B5347)；National Natural Science Foundation of China (NSFC) [61905243]; Scientific research project of Education Department of Jilin Province [JJKH20220992KJ]; Jilin Province Science & Technology Development Program Project in China [20200401071GX];

**Disclosures.** The authors declare no conflicts of interest.

**Data availability.** The data that supports the findings of this study are available within the article.